\documentstyle[12pt,aaspp4,tighten]{article}

\newcommand{\lsun}{log $L/L_{\odot}\,$}
\newcommand{\msun}{$M/M_{\odot}\,$}

\begin{document}

\title{METAL RICH RR LYRAE VARIABLES: II. THE PULSATIONAL SCENARIO}

\author{Giuseppe Bono}
\affil{Osservatorio Astronomico di Trieste, Via G.B. Tiepolo 11,
34131 Trieste, Italy; bono@oat.ts.astro.it}

\author{Filippina Caputo}
\affil{Osservatorio Astronomico di Capodimonte, Via Moiariello, 
16, 80131 Napoli, Italy; caputo@astrna.na.astro.it} 

\author{Santi Cassisi \altaffilmark{1}}
\affil{Osservatorio Astronomico di Teramo, Via M.Maggini 47, 
64100 Teramo, Italy; cassisi@astrte.te.astro.it} 

\author{Roberta Incerpi}
\affil{Dipartimento di Fisica, Univ. di Pisa, Piazza Torricelli 2,
56100 Pisa, Italy; roberta@astr1pi.difi.unipi.it}
\and

\author{Marcella Marconi}
\affil{Dipartimento di Fisica, Univ. di Pisa, Piazza Torricelli 2,
56100 Pisa, Italy; marcella@astr1pi.difi.unipi.it}

\altaffiltext{1}{Dipartimento di Fisica, Universit\`a de L'Aquila,
Via Vetoio, 67100 L'Aquila, Italy}

\date{ }

\begin{abstract} 

\noindent
We present a theoretical investigation
on the pulsational behavior of metal-rich RR Lyrae variables
over the range of evolutionary parameters suitable for stars 
with metallicities Z= 0.006, 0.01 and 0.02. With the addition of similar
results for metal-poor pulsators we discuss the theoretical pulsational 
scenario covering the metallicity range from Z=0.0001 to 0.02.

By connecting pulsational constraints to evolutionary prescriptions
for He burning stars we discuss the observed behavior
of the RR Lyrae population in the Galactic field.
We find that the distribution of field $ab$-type RR Lyrae stars
in the period-metallicity plane can be easily understood within the 
framework of the present theoretical scenario, suggesting that the 
Oosterhoff dichotomy also affects field variables. 

Theoretical predictions concerning the amplitude-period diagram
are discussed and compared with observational data. 
We find a general agreement for metal-poor ([Fe/H]$<$-1.4) RR Lyrae stars, 
whereas more metal-rich variables show amplitudes smaller than 
those predicted for pulsators originated from old, low-mass evolving stars. 
Alternatively, the agreement between theory and observations would 
require that a substantial fraction of metal-rich RR Lyrae variables 
in the Galactic field were younger than $\approx$ 2 Gyr.

The comparison between the pulsational behavior of RR Lyrae either 
in the Galactic 
field or in the Galactic bulge discloses the evidence that, at least as 
far as RR Lyrae variables are concerned, the metal-rich components of the 
bulge and of the field population appear quite similar.
We finally suggest that the peak at log P$\simeq$-0.55 in the
period frequency distribution of 
first overtone RR Lyrae stars in the Large Magellanic Cloud, 
considered as possible evidence for second overtone pulsators,
could be more simply taken as evidence of a metal-rich stellar population.
\end{abstract}

\noindent
{\em Subject headings:} Galaxy: stellar content -- stars: evolution 
-- stars: horizontal branch -- stars: oscillations 
-- stars: variables: RR Lyrae

\section{INTRODUCTION}

\noindent 
In a previous paper (Bono et al. 1996, hereinafter Paper I)
we investigated the theoretical predictions concerning  
metal-rich, low-mass evolutionary structures 
undergoing radial pulsational instabilities
during their central He burning phase. Summarizing the results,
we found that if the increase of He with metals is taken into account and 
if the stars are old enough (i.e. older than a few billion years), 
then He burning structures with metallicity $Z$ between 0.01 to 0.02 
should cross the region for radial instability at luminosities which 
range from \lsun$\simeq$1.52 to 1.54, with stellar masses from 
$M\simeq 0.55$ to 0.58$M_{\odot}$. 
However, in the same Paper I we showed that if younger He  
burning stars are pushed  into the instability region by 
a more efficient mass loss during the Red Giant Branch (RGB) phase, 
one could expect much lower luminosities together with much smaller 
masses. For ages close to 1 Gyr and by assuming a very efficient 
mass loss ($\approx 1.6 M_{\odot}$) we obtain as lower limits 
\lsun${\simeq}$1.1 for the ZAHB luminosity level and \msun${\simeq}$0.36 
for the stellar mass.

Taking into account such evolutionary  scenario, in this 
paper we approach the pulsational behavior of He burning models. 
The aim is to provide a theoretical scenario to be compared with 
observational data of metal-rich RR Lyrae stars observed  
both in the Galactic field and in the Galactic bulge.
Since the pulsational behavior depends not only on the 
structural parameters (stellar mass, luminosity and effective temperature) 
but, mainly at larger metallicities, also on the amount of metals 
(see Bono, Incerpi  \& Marconi 1996, hereinafter BIM), 
several {\it ad hoc} sequences of nonlinear, nonlocal and time-dependent 
convective models of RR Lyrae variables have been computed and discussed.

In the next section we investigate the region of the HR diagram
where  evolutionary stellar structures with solar metallicity ($Z$=0.02)
reveal to be unstable against radial pulsation, reporting the 
full results of nonlinear computations. Section 3 gives
similar results but for the cases $Z$=0.01 and $Z$=0.006. 
In Section 4 theoretical constraints are compared with observational 
data of RR Lyrae stars in the Galactic field, also with reference to 
previous results concerning metal-poor RR Lyrae stars, while 
Section 5 deals with variables in the Galactic bulge. The summary of 
the leading results of this investigation and some final remarks on RR 
Lyrae stars in the MACHO database close the paper.

\section{SOLAR METALLICITY RR LYRAE VARIABLES}

\noindent 
As a first step in approaching the pulsational scenario concerning metal-rich 
pulsators we investigate the location in the HR diagram of the 
regions where stellar envelope models become pulsationally unstable. 
Following the evolutionary results, this investigation has been 
performed by assuming fixed stellar mass (\msun=0.53) and chemical 
composition ($Z$=0.02, $Y$=0.28). 
The HR diagram has been explored with steps of 100 $K$ in the effective 
temperatures for  six choices of stellar luminosity, namely 
\lsun = 1.21, 1.41, 1.51, 1.61, 1.81 and 2.00.  As shown by the 
evolutionary results given in Paper I, this choice allows for a full 
coverage of the pulsational candidates of He burning structures with ages 
in the range 1 $\div$ 20 Gyr and takes into account both the ZAHB and the 
off-ZAHB evolutionary phases. The nonlinear theoretical approach 
adopted for this analysis has been already discussed in a previous series of 
papers (e.g. see Bono \& Stellingwerf 1994, hereinafter BS; 
Bono et al. 1996, hereinafter BCCM) and therefore it has not been discussed 
further.

\begin{table} 
\begin{tabular}{ccccc}
\multicolumn{5}{c}{TABLE 1} \\  
\multicolumn{5}{c}{Fundamental and first overtone boundaries.} \\ 
\tableline
\tableline
 \lsun &FOBE\tablenotemark{a}&FBE\tablenotemark{b}&FORE\tablenotemark{c}&FRE\tablenotemark{d} \\
        & (K)  & (K) & (K)  & (K) \\
\tableline
\multicolumn{5}{c}{$M=0.53 M_{\odot}$, $Y=0.28$, $Z=0.02$} \\  
\tableline
2.00 &  & 6650  &  & 4950 \\
1.81 & 6750 & 6750 & 6650 & 5350 \\
1.61 & 7050 & 6950 & 6150 & 5650 \\
1.51 & 7150 & 6850 & 6725 & 5850 \\
1.41 & 7350 & 6950 & 6550 & 5850 \\
1.21 & 7550 & 6850 & 6650 & 6150 \\
\tableline
\multicolumn{5}{c}{$M=0.58 M_{\odot}$, $Y=0.255$, $Z=0.01$} \\  
\tableline
1.65 & 7050 & 6950 & 6550 & 5650 \\
1.57 & 7150 & 7050 & 6550 & 5750 \\
1.51 & 7150 & 7050 & 6550 & 5850 \\
\tableline
\multicolumn{5}{c}{$M=0.58 M_{\odot}$, $Y=0.255$, $Z=0.006$} \\  
\tableline
1.55 & 7250 & 7150 & 6650 & 5850 \\
\tableline 
\end{tabular}
\tablenotetext{a}{Effective temperature of first overtone blue edge.}
\tablenotetext{b}{Effective temperature of fundamental blue edge.}
\tablenotetext{c}{Effective temperature of first overtone red edge.}
\tablenotetext{d}{Effective temperature of fundamental red edge.}
\end{table} 

Figure 1 reports the result of this investigation (solid lines), disclosing 
the regions in the HR diagram where the fundamental and/or the first 
overtone mode result unstable. Similar results but for metal-poor 
stars, as given by BCCM, are also shown (dashed lines). The dotted lines 
give the zero age horizontal branch (ZAHB) luminosity at log$T_e$=3.85 
for "old" structures (see Paper I) and for the labeled assumptions about
metallicity. 
Around this luminosity level, and for each assumed mass, 
moving from higher to lower effective temperatures the different lines 
show the first overtone blue edge (FOBE), the fundamental blue edge 
(FBE), the first overtone red edge (FORE), and the fundamental 
red edge (FRE). Luminosities and effective temperatures 
of the computed pulsational boundaries are listed in Table 1. 

Figure 1 shows that moving from extreme population 
II to "solar" pulsators the luminosity of the stars and the topology 
of the instability strip move in such 
a way that the range of effective temperatures where stars experience 
pulsational instability appears largely unchanged. As a 
matter of fact, mainly for the effect of the decreased mass, the solar
metallicity strip largely reproduces the topology of the metal-poor
case, but at smaller luminosities. 
According to such a topological transformation, in the solar case
we find that the upper luminosity level which allows for the occurrence 
of first overtone pulsators has now decreased down to $\log{L}\simeq$
1.8. Above this limit only fundamental pulsators are expected. 

Selected 
pulsating models for each assumed level of luminosity and for given values 
of the stellar effective temperature have been followed in time 
until the radial motion approaches its asymptotic amplitude. Tables 2 and 3 
in the Appendix give selected
quantities for all the computed models while Fig. 2 and Fig. 3
show an atlas of the bolometric light curves for both 
fundamental and first overtone pulsating modes. Comparison with
the light curves of metal-poor pulsators allows for some 
general comments. As far as the fundamental mode is concerned, the 
overall light
curve morphology found in the metal-poor case appears translated to
lower luminosity levels, including the occurrence 
of secondary features like "Bumps" and "Dips" through the pulsation cycle. 
However, for \lsun=1.81 the two bluest models
of the metal-rich series lack the "Dip" on the rising branch and present
a little "Bump" on the decreasing branch, a feature which is not found 
in any of the metal-poor models. Concerning both the instability strip 
topology and the light curve morphology as a whole it appears that 
metal-rich pulsators largely reproduce, although at lower luminosities, 
the theoretical scenario presented in BS and in BCCM for metal-poor RR 
Lyrae variables, with some minor variations which could be revealed only 
in very accurate photometric light curves or through a Fourier analysis
of the theoretical light curves.

First overtone light curves again  display properties similar to those of   
metal-poor pulsators, but shifted toward lower luminosity levels if 
compared to globular cluster pulsators. 
Again we find that at the lower luminosities, i.e. below
the predicted HB luminosity level, first overtone  models are expected 
to show an asymmetric light curve, resembling  the fundamental mode 
morphology. More in general, at the highest and lowest luminosity 
levels we find that the predicted light curves have a peculiar behavior 
which runs against the present observational scenario either of the 
fundamental pulsators (at \lsun=1.81) or of the first overtone ones 
(at \lsun=1.21). Such an evidence would thus constrain the luminosity 
of actual pulsators roughly to \lsun=  1.4$\div$1.6, in 
qualitative but relevant agreement with theoretical predictions.

As already discussed in BCCM, the 
pulsational investigation offers the relevant opportunity
of deriving theoretical predictions about the behavior of both
period and amplitude of the pulsation. The periods appear in fair 
agreement with the analytical dependence on luminosity,  
effective temperature, and mass value given for metal-poor stars in
BCCM. Specifically, for each given value of effective temperature and 
luminosity, 
the periods for first overtone pulsators do not show relevant differences
($\delta{\log{P}\approx{0.01}}$), 
whereas those of fundamental pulsators appear systematically larger
than analytical predictions by an amount of $\delta{\log{P}\approx{0.03}}$ . 

Figure 4 shows the dependence of the bolometric 
amplitudes on the effective temperature for the different assumptions 
on the luminosity level. We find the interesting evidence that around 
the expected luminosity of HB pulsators (namely, 
for \lsun in the range from 1.4 to 1.6)
the amplitudes of fundamental pulsators follow, with
good accuracy, the amplitude-temperature 
relation already found for metal-poor variables. More in detail, 
we find that for each given effective temperature the bolometric 
amplitude increases very slowly as the luminosity increases. Similarly, 
first overtone pulsators show the  characteristic "bell" shape
in the amplitude-temperature plane displayed by metal-poor models
even though in this case the maximum amplitude is almost independent of the 
luminosity level except for the sequence of models at \lsun=1.21.

At the two highest luminosity levels the fundamental amplitudes show an
anomalous behavior: for \lsun=1.81 they increase as the  
effective temperature decreases between 5800 and 5600 $K$, whereas for 
\lsun=2.0 they resemble the "bell" 
shape disclosed by first overtone amplitudes at lower luminosities. 
This behavior appears related to the peculiar topology of the instability 
strip at these large luminosities, where only the fundamental mode 
presents a stable nonlinear limit cycle. In such a case 
the quoted behaviors can be regarded as evidence that in this region
of the instability strip the pulsational amplitudes of fundamental 
pulsators attain vanishing values close to the instability boundaries.
In the region of the instability strip where both 
fundamental and first overtone present a stable limit cycle, first 
overtone pulsators attain vanishing amplitudes close to the high 
temperature edge (FOBE), whereas fundamental pulsators close to 
the low temperature edge (FRE).  

The theoretical period versus bolometric amplitude diagram for selected 
luminosity levels is reported in Fig. 5. As discussed in BCCM,
the topology of data shown in Fig. 5 is easily 
understood in terms of data in Fig. 4 when the relations  connecting
periods to luminosity and to effective temperature are taken into account. 
The dependence of  period on both luminosity and effective temperature 
removes the "degeneracy" of fundamental amplitudes with 
stellar luminosity shown in Fig. 4.

\section {THE CONNECTION WITH METAL-POOR RR LYRAE VARIABLES}

\noindent 
In the previous section we presented theoretical expectations
covering the pulsational behavior of "old" ($t>2.5$, with $t$ in Gyr) 
He burning stars with
solar metallicity. In this way we dealt with a theoretical scenario
which should be adequate for a metallicity regarded as a safe upper 
limit for RR Lyrae stars in the Galactic field. However, 
field RR Lyrae stars cover a wide range of metallicities,
reaching values as low as  [Fe/H] $\simeq$-2.2 or less, typical of the 
most metal-poor Galactic globular clusters. The theoretical
pulsational scenario for globular cluster RR Lyrae has been already 
presented in BCCM and it can be safely used also for approaching
the problem of field metal-poor pulsators. According to the discussion
given in the introduction, in this section we intend to 
fill the gap between solar metallicity and metal-poor pulsators 
by investigating the behavior of "mild" metal-rich pulsators with $Z$=0.01
and 0.006. In this way we will derive a theoretical scenario covering 
the pulsational properties of RR Lyrae stars over the whole range of 
metallicity from $Z$=0.0001 to  $Z$=0.02. 

For the case $Z$=0.01, from Paper I we assume the values $M=0.58M_{\odot}$ 
and Y=0.255 as suitable evolutionary parameters for old
He burning pulsators. Recalling that for similar pulsators one
expects a ZAHB luminosity level \lsun=1.54, 
we explored the pulsational stability  
for three luminosity levels which should cover the 
evolutionary behavior of the models, namely \lsun = 1.51, 1.57, 1.65. 
Concerning 
$Z$=0.006, we assume the same total mass and helium content with  
the ZAHB luminosity level \lsun = 1.55. The full results are 
listed in Tables 4, 5 and 6 in the Appendix, while luminosities and 
effective temperatures of the computed pulsational boundaries are 
presented in Table 1.

Figure 6 shows the instability strip with $Z$=0.01 
(solid lines) together with the results 
with $Z$=0.006 (asterisks). In order to point out the 
dependence of the instability edges on the metal content, in this 
figure the instability strip for the solar case (dashed lines) 
is also shown. 
We find  that near the ZAHB luminosity level the boundaries of the 
instability strip become slightly redder as the metallicity increases, 
i.e. at fixed helium content, and passing from $Z=0.006$ to $Z=0.01$. 
However, as soon as an increase of both helium and 
metals is taken into account, the location of the boundaries is  
only marginally affected and the 
major difference seems to be connected with the topology of the
region between FBE and FORE, the so-called "OR" region.

Figures 7a-b show the atlas of theoretical light curves computed with 
$Z$=0.01 and the three luminosity levels, while Fig. 8
refers to the sequence of models with  $Z$=0.006. 
Again we find that theoretical light curves appear in 
good qualitative agreement with the observational scenario. Moreover,  
we find that the periods of  
first overtone pulsators follow the analytical relation 
given in BCCM ($\delta{\log{P}\approx{0.01}}$),  while the periods of 
fundamental pulsators are systematically larger by 
$\delta{\log{P}\approx{0.02}}$ in comparison with the values provided by the 
quoted relations. 

Figure 9  shows the relation between bolometric amplitude 
and effective temperature for
the $Z$=0.01 (solid lines) and $Z$=0.006 (dotted lines) 
pulsating models. Data plotted in this figure clearly display that the 
bolometric amplitudes are correlated with effective temperatures, but 
here with a very slow dependence on the assumed luminosity. 
Moreover, the same figure shows that for each given temperature the two 
sequences of models at $Z$=0.01 and 0.006 are characterized by bolometric 
amplitudes moderately larger if compared with the amplitudes obtained 
by adopting roughly the same luminosity but a solar metal content 
(dashed lines). Conclusions in a recent paper by BIM state that, while 
all physical parameters are constant,   
the fundamental amplitudes slowly increase with increasing $Z$. 
Therefore the opposite behavior found in Fig. 9 could be due to the 
increase of the helium content connected with the increase of metals.
As a matter of fact, the comparison between metal-poor pulsators 
with $Y$=0.30 (BS) and $Y$=0.24 (BCCM) suggests that at 
fixed luminosity and effective temperature the bolometric amplitude 
decreases if helium increases. Moreover, 
some models computed with $Y$=0.31, $Z$=0.02, \lsun=1.51 and 
0.53$M_{\odot}$ (plotted as asterisks in Fig. 9) actually show smaller 
amplitudes than similar models constructed by adopting $Y$=0.28. 
     
Theoretical predictions concerning the period-bolometric amplitude
relation for the cases $Z$=0.01 (solid lines) and $Z$=0.006 (dotted lines)
are finally given in Fig. 10, in comparison with the results (dashed lines) 
for mild metal-poor pulsators with $Y$=0.24, $Z$=0.001 (BCCM).
The models plotted in this figure have been computed by adopting the same 
mass value and therefore it allows for the analysis of the dependence 
of bolometric amplitudes on chemical composition. 
A glance at the curves plotted in this figure show that fundamental 
amplitudes at the ZAHB luminosity level slightly decrease as the metal 
and the helium contents increase. On the other hand, first overtone 
amplitudes do not show a substantial dependence on the chemical 
composition and the main difference is connected with the change 
in the period range.

\section{RR LYRAE VARIABLES IN THE GALACTIC FIELD}

\noindent 
The pulsational theoretical scenario presented in the previous
sections can now be connected with the evolutionary constraints
presented in Paper I in order to produce theoretical predictions about
the expected behavior of metal-rich RR Lyrae pulsators.
As a first step, we investigate the well known finding (Preston 1959) 
for which field metal-rich RR Lyrae stars are characterized by shorter 
periods when compared with metal-poor variables. 

For a  discussion of this point, we can rely on both the theoretical results 
concerning the location in the HR diagram of the blue boundaries for 
radial instability, derived from pulsational models, and the predictions 
concerning stellar masses and luminosities at those  boundaries derived 
from evolutionary models. In this framework predictions about the 
fundamental and first overtone minimum periods can be obtained for the 
different assumptions on metallicity, as reported in Fig. 11. 

The theoretical shortest period for 
first overtone pulsators is calculated at the intersection 
between ZAHB (see Table 6 in Paper I) and first overtone blue edge. 
Whereas, for the minimum fundamental period, we assume that the transition
from fundamental to first overtone pulsators occurs close to the 
fundamental blue edge or to the first overtone red edge. On this basis 
the region in Fig. 11 located  between the FBE and FORE lines 
represents the "OR" region, i.e. the region of the instability strip  
where a variable presents a stable limit cycle in both fundamental and  
first overtone modes. It is worth noting that the theoretical 
scenario was obtained by assuming, for all metallicities, "old" He burning
pulsators. Moreover, for $Z< 0.006$ (solid lines) we adopted 
$Y_{MS}$=0.23, while for $Z\ge 0.006$ (dashed lines) an enrichment ratio 
between helium and heavy elements $\Delta Y_{MS}$/$\Delta Z\approx$3 
was taken into account.

Figure 12 presents the comparison between the theoretical predictions 
of Fig. 11 for fundamental pulsators  and the observational data provided 
by Blanco (1992, hereinafter VBL92) for field {\it ab}-type RR Lyrae 
variables. The metallicities, based on Butler (1975) $\Delta {S}$
index, have been obtained by adopting the relation provided by 
Suntzeff, Kinman \& Kraft (1991), which, in turn, relies on the 
Zinn \& West (1984) metallicity scale. 
Open circles are RR Lyrae stars with uncertain or unknown
blue amplitude, while asterisks refer to variables with unreliable 
$\Delta {S}$ index. The error-bar refers to the different calibrations of 
$\Delta {S}$ index as a function of  [Fe/H] found in the literature 
(Butler 1975; Suntzeff et al. 1991; VBL92; Jurcsik 1995; 
Fernley \& Barnes 1996).  

Figure 12 reveals the interesting evidence that
the present pulsational-evolutionary scenario appears in fair 
agreement with RR Lyrae stars  in the Galactic field. 
As for the metal-poor variables ([Fe/H]$<$-1.4), they appear to 
reproduce the exhaustively discussed behavior of RR Lyrae variables 
belonging to Galactic globular clusters, with evidence for the occurrence  
of the Oosterhoff  phenomenon. As a matter of fact,
among the most metal-poor stars we recognize the lack of fundamental 
pulsators in the "OR" region and hence the larger values of the mean period
which characterize Oosterhoff type II (OoII) pulsators 
(see BCCM and references therein). By increasing the metallicity,
the "OR" region begins to be filled and the mean period decreases, with 
the pulsational behavior moving toward typical Oosterhoff type I (OoI)  
pulsators. 

Figure 12 also shows that for pulsators with  
[Fe/H]$>$-1.4 the predicted minimum fundamental period, as 
attained at the fundamental blue edge, decreases with increasing Z and 
reaches, in agreement with observations, $\log{P}=-0.45$ at Z$\approx$0.006.
At larger metallicities  the predicted FBE periods
appear consistent with the lower envelope of the observed distribution, with 
the exception of few stars and, in particular, TV Lib. Similar 
{\it stragglers} can be understood only assuming a lower age and therefore 
lower luminosities and smaller periods. This seems almost mandatory to 
account for the deviant star TV Lib.
However, it should be noted that  metal-rich variables could be 
evolved stars originated in the low temperature side of the instability strip 
and that the corresponding evolutionary tracks could be located at 
effective temperatures lower than FBE. In such a scenario, 
the predicted minimum period for fundamental pulsators 
should be larger than the one attained by the variable at FBE and 
consequently young metal-rich variables could be hidden
in the sequence of "old pulsators".

In order to evaluate the effects of observational errors and/or different 
metallicity calibrations, in Fig. 13 we present the sample of field stars 
recently analyzed by Layden (1994, 1995, hereinafter LA95) 
where [Fe/H]$_K$ is based on the strength of the CaII K line. 
Also for this sample the agreement between theoretical 
predictions and observed data is satisfactory, as well as for the 
Kemper (1982, hereinafter KE82) sample of field {\it c}-type variables 
(Fig. 14).

A much less satisfactory agreement is found when the comparison deals 
with  pulsational amplitudes. Figure 15 discloses the predicted Bailey 
diagram (blue amplitude {\it vs} period) for "old" pulsators at the ZAHB 
luminosity levels derived for different assumptions about stellar 
metallicities (for the values of stellar mass and luminosity see Table 6 
in Paper I) and with the bolometric amplitudes transformed into blue 
amplitudes by means of Kurucz (1992) atmosphere models. 

The bottom panel shows the predicted behavior with $Z$=0.0001 (dotted line), 
$Z$=0.001 (dashed line), and $Z$=0.006 (solid line), while the upper panel 
presents the comparison between the $Z$=0.006 case and the results for 
$Z$=0.01 (dashed line) and $Z$=0.02 (dotted line). 
As a whole, we derive a fair independence of the amount of metal over 
the range $Z=0.0001\div{0.02}$, thus excluding the occurrence 
of different sequences as functions of metallicity. 
However, it is worth underlining that under the 
hypothesis of young ($t\approx$1 Gyr) metal-rich pulsators 
the predicted period-amplitude sequence would move toward 
smaller periods, as shown by the dashed-dotted line for the 
case $Z$=0.02, \lsun=1.41. 

Comparison with observational data is given in Fig.s 16a-b. 
The bottom panel of Fig. 16a shows the Layden variables with [Fe/H]$_K <$ -1.4  
together with the predicted behavior of fundamental 
pulsators with $Z$=0.0001, $M=0.65M_{\odot}$, 
and \lsun=1.61 and 1.72. In the upper panel  
the variables with [Fe/H]$_K$ in the range from -0.6 to -1.4 
are plotted as open circles, while the solid line shows the 
predicted behavior of fundamental pulsators with $Z$=0.006, $M=0.58M_{\odot}$,
and \lsun=1.55. Figure 16b refers to metal-rich 
RR Lyrae stars ([Fe/H]$_K>$-0.6) in comparison with 
the predicted behavior of "old" 
and "young" ZAHB pulsators with $Z$=0.02. As a whole, 
there is  a fair agreement for the metal-poor variables, whereas the 
behavior of metal-intermediate and metal-rich components seems 
too discrepant to be accepted. Again, it should be noted that 
predicted amplitudes for "young" solar metallicity 
pulsators easily match observations (dashed-dotted line in Fig. 16b).
We conclude that either the theory overestimates the amplitude of metal-rich
RR Lyrae variables or their behavior in the Bailey diagram reveals 
the large occurrence of "young pulsators".

\section{RR LYRAE VARIABLES IN THE GALACTIC BULGE}

\noindent 
The pulsational scenario discussed in the previous sections  can 
be also compared with the pulsational behavior of RR Lyrae stars 
in the Galactic bulge. Baade (1951) observing in the so-called Baade's 
Window \footnote{It is interesting to note that Baade (1958) called this 
central region of the Galaxy {\em van Tulder's pole.}} 
(BW, centered on $l = 1^o$.0, $b=-3^o$.9) first discovered 
RR Lyrae stars in the Galactic bulge, using the magnitudes of these 
variables to estimate the distance to the Galactic center. 
With the improvement of photometric techniques 
further variables  were discovered in the BW (e.g. Oort \& Plaut 
1975; Blanco \& Blanco 1985; Walker \& Mack 1986), causing 
increasing attention to be devoted to this population of variable stars. 

Blanco (1984, hereinafter BBL84) investigated a sample of 77 
RR Lyrae variables in the BW suggesting that, in spite of a significant 
range of metallicity covered by these stars, most of them are relatively 
more metal-rich than stars in the Galactic globular cluster M3. 
The range of magnitudes appeared to be quite narrow, 
with important implications for the controversy about 
the slope of the RR Lyrae magnitude-metallicity relation.

Figure 17 shows the distribution of periods with metallicity, as
derived from $\Delta{S}$ values listed by Walker \& Terndrup (1991) 
and calibration by VBL92 and Suntzeff et al. (1991).  
It appears in fair agreement with theoretical 
predictions as well as with the behavior of field RR Lyrae stars with 
similar metallicities. Moreover, due to the moderately large metal abundance,
it appears that the transition between fundamental and 
first overtone pulsators occurs close to the fundamental blue
boundary, with no evidence for the OoII behavior.

Figure 18 shows the period-amplitude diagram
for BW {\it ab}-type RR Lyrae stars (lower panel) in comparison 
with  predictions for "old" (dotted line) and "young" 
solar metallicity pulsators at \lsun=1.41  
(dashed-dotted line). Open circles refer to 
variables with [Fe/H]$\ge$-1.0. Also in this case we find a sort 
of discrepancy between theory concerning old stars and observations, with the
conclusions already discussed at the end of the previous section. 

The behavior of first overtone pulsators (Fig. 19) 
adds to this scenario only the 
evidence that a metallicity $Z\approx$0.01 is almost suitable for 
observations and that a component of OoII variables is not 
present in the BW, as shown by the lack of the decreasing  
branch characterizing this type of pulsators (see BCCM).
This confirms that the bulge metallicity does not reach 
the low values which are typical of OoII globular clusters.  
Moreover, first overtone periods group around log P$\simeq$-0.55,
(P$\simeq$0.3 days) \footnote{This peculiarity in the period 
distribution of RR Lyrae variables in the central region of our 
Galaxy was originally pointed out by Baade (1958) long before  
the metal abundances of these objects had been estimated !}, 
in agreement with the expectation for metal-rich stars given in 
the previous section. 
A similar secondary peak in the period distribution of first overtones  
is also present in some Dwarf Spheroidal Galaxies belonging to the Local 
Group like Ursa Minor (Nemec, Wehlau, \& de Oliveira 1988) and Sculptor 
(Kaluzny et al. 1995, and references therein) and is further stressed 
by recent data collected by the OGLE project 
(Udalski et al. 1994, 1995a,b) for pulsators in BW. 

Surveys in  bulge windows other than the BW are really important,
since interstellar absorption in BW is rather uneven and extreme 
crowding affects the photometric accuracy. A survey of a new Galactic bulge
window ($l=0.6$, $b=-5.5$), which is remarkably uniform in absorption
and not as seriously affected by crowding as the BW, was carried out by 
Blanco (1992, hereinafter BBL92) who discovered and studied 112 new RR Lyrae 
variables. According to the quoted author, the comparison between
the mean periods of RRab and RRc in this new window and those in the BW
possibly shows a difference which seems to suggest a metallicity 
gradient between the two windows. 

Data for fundamental pulsators in the 
new quoted window are reported in the upper panel of 
Figure 18. As a whole, the pulsational behavior of variables   
in both windows appears rather homogeneous, not giving clear evidence
for the suggested metallicity gradient. Moreover, the comparison with 
Figs. 16a-b suggests that, at least as far as RR Lyrae stars 
are concerned, the metal-rich component of the bulge population 
does not differ from similar stars present in the Galactic field.

\section {SUMMARY AND CONCLUSIONS}

\noindent 
In a homogeneous theoretical context we developed both evolutionary
and pulsational properties of low-mass helium burning stars with  
chemical compositions typical of field RR Lyrae variables. The evolutionary 
calculations, based on canonical HB models, have been already 
discussed in a previous companion paper (Paper I). The pulsation 
characteristics and the modal stability of hydrodynamical envelope models 
have been derived by adopting a nonlinear, nonlocal and time-dependent 
convective approach. For each given chemical composition the sequences of 
pulsating models were constructed by using both the stellar masses and 
the luminosity levels predicted by HB evolutionary tracks. 
On the basis of this thorough theoretical investigation and of the comparison 
with the presently available observational scenario concerning RR Lyrae 
variables in the Galactic field and in the Galactic bulge, we can draw 
the following conclusions:

{\em a.} At fixed stellar mass and chemical composition the shape of the 
light curve
of metal-rich pulsators shows a dependence on both effective temperature and 
luminosity quite similar to that of metal-poor RR Lyrae models constructed 
by BS and BCCM. Besides some minor differences connected with the 
appearance of secondary features along the light curves, the instability 
strip topology of metal-rich pulsators (fundamental and first overtone) 
is shifted toward luminosities lower than those of metal-poor pulsators.   

{\em b.} On the basis of few selected models computed by adopting fixed stellar 
mass, luminosity level, and metal content but different helium contents
(Y=0.28, 0.31) we find that the bolometric amplitude of fundamental 
pulsators decreases as helium increases. 
 
{\em c.} The comparison in the period-metallicity plane between Galactic field 
fundamental variables collected by VBL92 and theoretical predictions  
shows, within the range of observational uncertainties connected with 
the metallicity calibration of the $\Delta S$ index, a satisfactory 
agreement over a wide 
metallicity range (2 dex). At the same time there is evidence of the 
appearance of the Oosterhoff dichotomy among field variables. In fact,
in the "OR" region there is substantial lack of metal-poor ([Fe/H]$<$ -1.7) 
fundamental pulsators, which, in turn, implies that this group 
of pulsators resembles OoII variables, whereas for higher metal contents
field variables follow the pulsation characteristics of OoI variables. 
The agreement between theory and observations is still 
satisfactory if we take into account the sample of field fundamental 
variables recently provided by Layden (1995). The outcome is similar 
for the sample of first overtone field variables collected by 
Kemper (1982). 

{\em d.} The results obtained from the comparison in the Bailey plane 
-blue amplitude versus period- between theoretical predictions and 
observations are much less straightforward. As a matter of fact, 
the B amplitudes of metal-poor ([Fe/H]$<$-1.4) pulsators are in fair 
agreement with theoretical amplitudes obtained by assuming the ZAHB 
luminosity levels of "old" pulsators. On the contrary, the predicted 
amplitudes for both metal-intermediate (-1.4 $<$[Fe/H]$<$-0.6) and 
metal-rich ([Fe/H]$>$-0.6) pulsators are systematically larger in 
comparison with the observed ones. However, in this context it is worth 
noting that the predicted amplitudes for "young" ($t\approx$ 1 Gyr) 
solar metallicity pulsators provide a satisfactory fit for observational 
data. 

{\em e.} The comparison between theoretical predictions and observational 
data was also extended to the RR Lyrae variables belonging to the Galactic 
bulge. On the basis of both observational data provided by BBL84 and BBL92  
and the theoretical scenario previously outlined,  
the comparison in the period-metallicity plane is quite satisfactory, 
whereas the Bailey diagram presents the same discrepancy disclosed 
by field RR Lyrae variables. The assumption of "young" pulsators could 
provide, for this sample of variables, a good fit for the pulsational  
amplitudes of both fundamental and first overtone pulsators. 

{\em f.} The nonlinear fundamental periods of the present survey are 
systematically larger ($\delta log P\approx 0.03$) in comparison with 
the periods given by the analytical relation suggested by BCCM. 
This discrepancy clearly suggests that when moving from metal-poor 
to metal-rich pulsators the nonlinear periods show a nonnegligible 
dependence on both helium and/or metal contents. 

{\em g.} As a final point, the pulsational scenario discussed in this 
paper allows for an approach to the results of the MACHO survey of 
RR Lyrae in the
Large Magellanic Cloud presented by Alcock et al. (1993, 1995, 1996).
As for the period-amplitude diagram, we note that the discussion 
in that paper was partially based on the assumption of different sequences 
characterizing OoI and  OoII fundamental pulsators in the Bailey diagram, 
an occurrence already challenged both on theoretical (BCCM) and observational
(Brocato, Castellani \& Ripepi 1996) grounds and confirmed here. 
As a more relevant point, it has been suggested that the peak in the 
period frequency distribution of first overtone pulsators located at 
log P$\simeq$ -0.55 is the possible evidence of second overtone RR Lyrae 
pulsators. Here we advance the hypothesis that such a peak could be taken 
as evidence of a metal-rich population, in agreement with the behavior 
we already found for both Galactic field and Galactic bulge variables. 

\noindent 
It is a pleasure to thank V. Castellani for many useful and enlightening 
insights on the different outcomes of this investigation and for a 
critical reading of an early draft of this manuscript. 

\appendix

\section{NONLINEAR MODEL SUMMARY: PULSATIONAL AMPLITUDES} 

\noindent 
Although several and thorough investigations have been devoted to the 
nonlinear systematic properties of RR Lyrae variables in Galactic globular 
clusters since Christy's seminal paper (1966),  
we still lack detailed analysis of the pulsation behavior of 
metal-rich RR Lyrae variables observed in the Galactic field and in  
the Galactic bulge. 
With the aim to fill this gap, in this Appendix we report the results 
of the nonlinear survey of RR Lyrae models we computed. The sequences of 
envelope models have
been constructed by assuming different stellar masses and chemical 
compositions. Moreover, in order to provide a comprehensive theoretical 
framework for the pulsation properties inside the instability strip, we 
adopted a wide range of luminosity levels and effective temperatures. 
Since we are also interested in the modal behavior of envelope models,   
the stability analysis was performed for both fundamental and first 
overtone modes. With the primary purpose of locating the exact 
instability boundaries of these modes the structure equations have 
been followed in time until the radial motions approach their limiting 
amplitude. The reader interested to the methods and the assumptions 
adopted for constructing the pulsation models is referred to BS, BCCM,
and BIM. 

\noindent 
The overall stability results of this survey have been already discussed 
in previous sections. In Tables 2 to 6 are reported the pulsational 
amplitudes to be compared with actual RR Lyrae observational properties.
Similar data but referred to metal-poor pulsators have been already 
discussed by BCCM. Each table gives -left to right- 
1) logarithmic luminosity level; 
2) nonlinear period (days); 3) effective temperature (K); 
4) fractional radius variation, 
$\Delta R/R_{ph}=(R^{max}\,-\,R^{min})/R_{ph}$ where $R_{ph} $ is the 
photospheric radius; 
5) radial velocity amplitude (km/sec), $\Delta u=u^{max}\,-\,u^{min}$; 
6) bolometric amplitude (mag), $\Delta M_{bol}=M_{bol}^{max}\,-\,M_{bol}^{min}$;
amplitude of logarithmic surface gravity, 
$\Delta log g=log g^{max}\,-\,log g^{min}$, 7) static, $g_s=GM/R_{ph}^2$, 
and 8) effective, $g_{eff}=GM/R_{ph}^2\,+\,du/dt$; 
9) surface temperature variation, $\Delta T=T^{max}\,-\,T^{min})$ where $T$ 
is the temperature of the outer boundary;
10) effective temperature variation, 
$\Delta T_e=T_e^{max}\,-\,T_e^{min})$ where $T_e$ is derived from the 
surface luminosity and radius variations along a full cycle. 

\noindent 
The temperature values listed in columns 9 and 10 have been rounded up to
the nearest 50 K. In these tables we have reported only models which 
approach a stable nonlinear limit cycle. The models which, during the 
integration, experience a mode switch (fundamental toward first overtone 
or vice versa), or mixed-mode features (two or more radial modes are 
contemporary excited) are marked by upper letters located close to the 
value of the effective temperature. Close to the instability boundaries
we adopted an effective temperature step of 100 K. Therefore at fixed 
luminosity level the blue (red) edge of each mode can be easily derived 
by increasing (decreasing) by 50 K the effective temperature of the 
first (last) stable model reported in the following tables.

\pagebreak

\pagebreak
\section{Figure Captions}

\vspace*{3mm} \noindent {\bf Fig. 1.} The HR diagram location of 
the instability strip for fundamental and first overtone modes 
computed at fixed stellar mass (\msun=0.53) and solar metallicity 
(solid lines) as compared with the instability strip for metal-poor 
stars (dashed lines). The dotted lines show the ZAHB luminosity levels
for the labeled metal abundances.

\vspace*{3mm} \noindent {\bf Fig. 2.} Theoretical light curves of 
solar metallicity fundamental pulsators over two consecutive periods. 
The labels show the luminosity levels and the effective temperatures.

\vspace*{3mm} \noindent {\bf Fig. 3.} Same as in Fig. 2, but the light
curves are referred to first overtone pulsators.

\vspace*{3mm} \noindent {\bf Fig. 4.} Fundamental (solid lines) and 
first overtone (dashed lines) bolometric amplitudes versus effective 
temperatures for solar metallicity RR Lyrae variables. The luminosity 
levels are indicated by different symbols. 

\vspace*{3mm} \noindent {\bf Fig. 5.} Fundamental (solid lines) and 
first overtone (dashed lines) bolometric amplitudes versus periods 
for solar metallicity RR Lyrae variables. The luminosity 
levels are indicated by different symbols. 

\vspace*{3mm} \noindent {\bf Fig. 6.} Comparison between the instability
strips of models with $M=0.58 M_{\odot}$, $Y$=0.255, $Z$=0.01 (solid lines) 
and the solar metallicity models (dashed lines). The asterisks mark the 
instability boundaries for the sequence of models with $M=0.58 M_{\odot}$, 
$Y$=0.255, and $Z$=0.006. The dotted line shows the ZAHB luminosity level 
for $Z$=0.01 models. 

\vspace*{3mm} \noindent {\bf Fig. 7a.} Theoretical light curves over two 
consecutive periods  for fundamental pulsators with $M=0.58M_{\odot}$, 
$Y$=0.255, and $Z$=0.01.  
The labels show the luminosity levels and the effective temperatures.

\vspace*{3mm} \noindent {\bf Fig. 7b.} Same as in Fig. 7a, but the light
curves are referred to first overtone pulsators.

\vspace*{3mm} \noindent {\bf Fig. 8. } Same as in Fig. 7a, but the light
curves are referred to fundamental and first overtone pulsators
with $M=0.58 M_{\odot}$, $Y$=0.255, and $Z$=0.006.

\vspace*{3mm} \noindent {\bf Fig. 9.}  The relation between bolometric 
amplitude and effective temperature for fundamental and first overtone 
pulsators with $M=0.58 M_{\odot}$, $Y$=0.255, and $Z$=0.01 (solid lines) 
as compared with similar results but for different chemical compositions, 
namely  $Y$=0.28, $Z$=0.02 (dashed lines) and $Y$=0.255, $Z$=0.006 
(dotted lines). The asterisks mark the bolometric amplitudes of fundamental 
pulsators computed by adopting $Y$=0.31 and $Z$=0.02.

\vspace*{3mm} \noindent {\bf Fig. 10.}  The relation between
bolometric amplitudes and periods at fixed stellar mass $M=0.58 M_{\odot}$ 
and different chemical compositions. The labels show the adopted chemical 
compositions and the luminosity levels. 

\vspace*{3mm} \noindent {\bf Fig. 11.}  Theoretical
expectations about the minimum period as a function
of stellar metallicity. The minimum fundamental period 
is evaluated assuming that the transition
between fundamental and first overtone pulsators occurs either
at the fundamental blue edge (FBE) or at the first overtone red edge
(FORE). The minimum  first overtone period is evaluated at the first 
overtone blue edge (FOBE). Solid lines refer to $Y_{MS}$=0.23, whereas 
dashed lines refer to an original He which increases with increasing 
Z (see text for further details).

\vspace*{3mm} \noindent {\bf Fig. 12.} Comparison between predicted 
minimum fundamental periods and {\it ab}-type RR Lyrae stars in the 
Galactic field. Observational data from VBL92. Open circles refer 
to RR Lyrae with uncertain or unknown blue amplitudes, while asterisks 
display variables with unreliable $\Delta{S}$ index.

\vspace*{3mm} \noindent {\bf Fig. 13.} Same as in Fig. 12 but with 
observational data from Layden (1994) and LA95. 

\vspace*{3mm} \noindent {\bf Fig. 14.} Comparison between predicted
minimum first overtone periods and {\it c}-type RR Lyrae stars in the 
Galactic field. Observational data from KE82.
    
\vspace*{3mm} \noindent {\bf Fig. 15.} ({\em Bottom:}) The predicted Bailey 
diagram for "old" ZAHB pulsators with $Z$=0.0001 (dotted line), 
$Z$=0.001 (dashed line), and $Z$=0.006 (solid line). 
({\em Top:}) The predicted Bailey diagram for "old" ZAHB pulsators with 
$Z$=0.02 (dotted line), $Z$=0.01 (dashed line), and $Z$=0.006 (solid line). 
The dotted-dashed line refers to "young" pulsators with $Z$=0.02.

\vspace*{3mm} \noindent {\bf Fig. 16a.} Comparison of predicted blue 
amplitudes and periods with observational data for selected field 
{\it ab}-type RR Lyrae variables from Layden (1994) and LA95. 
Full circles show variables with [Fe/H]$<$-1.4, while open circles 
refer to variables with [Fe/H] in the range from  -1.4 to -0.6. 
The two dashed lines refer to ZAHB pulsators with $Z$=0.001 and two 
different choices of luminosity (\lsun=1.61, 1.72), while the solid 
line refers to ZAHB pulsators with $Z$=0.006. 
 
\vspace*{3mm} \noindent {\bf Fig. 16b.} Same as in Fig. 16a but for 
field variables with [Fe/H]$>$-0.6 and with the predicted blue 
amplitudes and periods referred to "old" (dotted line) and "young" 
(dashed-dotted line) ZAHB pulsators with $Z$=0.02.

\vspace*{3mm} \noindent {\bf Fig. 17.} Same as in Fig. 12 and Fig. 14,  
but for {\it ab}-type and {\it c}-type RR Lyrae stars in the 
Galactic bulge as given by BBL84. 

\vspace*{3mm} \noindent {\bf Fig. 18.} Same as in Fig. 16b, but for 
fundamental pulsators in the BW (bottom panel: BBL84) with [Fe/H]$\ge$-1.0 
(open circles) and [Fe/H]$<$-1.0 (full circles) and in the new Galactic 
bulge window discussed in the text (top panel: BBL92).

\vspace*{3mm} \noindent {\bf Fig. 19.} The  location in the Bailey
diagram of {\it c}-type RR Lyrae variables in the BW (BBL84) in comparison 
with predicted blue amplitudes and periods of first overtone pulsators 
with $Z$=0.01 and \lsun=1.51 (dashed line), 1.57 (dashed-dotted line), 
and 1.65 (dotted line).

\end{document}